\documentclass[preprint,3p]{elsarticle}
\usepackage{graphicx}
\usepackage{epstopdf}
\usepackage{graphics}
\usepackage{enumitem}
\usepackage{amssymb}
\usepackage{color}

\journal{Applied Energy}

\begin{document}

\begin{frontmatter}

\title{Influence of Thickness and Contact Area on the Performance of PDMS-Based Triboelectric Nanogenerators}

\author{A. Gomes\textsuperscript{}}
\author{C. Rodrigues\textsuperscript{}}
\author{A. M. Pereira\textsuperscript{}}
\author{J. Ventura\textsuperscript{}}

\address{\textsuperscript{}IFIMUP and IN-Institute of Nanoscience and Nanotechnology, Departamento de F\'isica e Astronomia da Faculdade de Ci\^encias da Universidade do Porto, Rua do Campo Alegre 687, 4169-007 Porto, Portugal}

\begin{abstract}
Triboelectric nanogenerators (TENGs) are an emerging mechanical energy harvesting technology that was recently demonstrated. Due to their flexibility, they can be fabricated in various configurations and consequently have a large number of applications. Here, we present a study on the influence of the thickness of the triboelectric layer and of the contact surface area between two triboelectrical materials on the electric signals generated by a TENG.
Using the PDMS-Nylon tribo-pair, and varying the thickness of the PDMS layer, we demonstrate that the generated voltage decreases with increasing thickness.
However, the maximum generated current presents an inverse behaviour, increasing with increasing PDMS thickness. The maximum output power initially increases with increasing PDMS thickness up to 32 $\mu$m, followed by a sharp decrease. Using the same tribo-pair (but now with a constant PDMS thickness), we verified that increasing the contact surface area between the two tribo-materials increases the electrical signals generated from the triboelectric effect.

\end{abstract}


\end{frontmatter}

\section{Introduction}
The most recent energy harvesting technology comes from the triboelectric effect \cite{Fan2012,Wang2014,Wen2014,Wang2015}. Triboelectric nanogenerators (TENGs) are based on the conjunction of triboelectrification and electrostatic induction in which a material becomes electrically charged after it comes into contact with another material through friction \cite{doi:10.1080/00018738000101466,Niu2014a,Lin2016,Fan2016}.
The discovery of TENGs opened a new field for materials scientists to fabricate nanogenerators that convert mechanical energy at high efficiencies \cite{Fan2012,Wang2014,Zhang2016,Khan2016,Rodrigues2016} and that are easy to integrate \cite{CSSC:CSSC201403481}.
This type of nanogenerators have innumerous advantages, such as flexibility, environmental friendliness, versatility and extremely high output voltages.

TENGs can have different configurations, depending on the way the two triboelectric materials come into contact, leading to four operation modes: vertical contact-separation, lateral-sliding, single-electrode and freestanding triboelectric-layer \cite{Wang2014,Niu2014a,Rodrigues2016,Niu2014,Niu2015,Niu2013a,Cao2016}.
Theoretical TENGs studies \cite{CSSC:CSSC201403481,Niu2013,Wang2016,Yang2016} found that, as first approximation, the output voltage (V) of a dielectric-to-dielectric TENG in the contact-mode is given by \cite{Niu2014a,Niu2014b}:
\begin{equation} \label{eq:1}
 V=-\frac{Q}{A\varepsilon_{0}}[d_{0}+x(t)]+\frac{\sigma x(t)}{\varepsilon_{0}}
\end{equation}
where $A$ is the triboelectric surface area, Q is the transferred charge, $x(t)$ is the time-dependence distance between the two triboelectric layers, $\varepsilon_{0}$ is the permittivity of free space and $d_{0}$ is the effective dielectric thickness given by $\frac{d_{1}}{\varepsilon_{r1}}$ + $\frac{d_{2}}{\varepsilon_{r2}}$ (where $d_{1}$ and $d_{2}$ are the thicknesses of the two dielectric materials and $\varepsilon_{r1}$ and $\varepsilon_{r2}$ are the relative dielectric constants of dielectrics 1 and 2). In the open-circuit case, there is no charge transfer ($Q=0$), so the open-circuit voltage ($V_{OC}$) is given by:

\begin{equation} \label{eq:2}
 V_{OC}=\frac{\sigma x(t)}{\varepsilon_{0}}
\end{equation}

The short-circuit current ($I_{SC}$) generated by a TENG is proportional to the generated triboelectric charge density ($\sigma$), the surface of the electrode ($A$), and the speed of the relative mechanical movement [$v(t)$] and inversely proportional to the square of the distance between the electrodes:

\begin{equation} \label{eq:3}
 I_{SC}=\frac{\sigma Ad_{0}v(t)}{[d_{0}+x(t)]^2}
\end{equation}

In the construction of triboelectric nanogenerators it is necessary to consider the main factors behind high performance, particularly the choice of triboelectric materials.
For this aim, one uses the \textit{Triboelectric Series}, where the triboelectric materials are ordered according to their polarity (ability of a material to gain/lose electrons) \cite{Wang2016,Gooding2014,Burgo2016}.
Materials such as glass or Nylon are positive tribo-materials and tend to lose electrons when coming into contact with negative charge tendency materials [e.g. Poly(tetrafluoroethylene) (PTFE) or Poly(dimethylsiloxane) (PDMS)], that have a tendency to gain electrons \cite{Wang2016,Gooding2014,Diaz2004}.
For this study, we choose PDMS and Nylon as the two triboelectrical materials, due to their opposite position in the triboelectric series \cite{Gooding2014,Diaz2004}.
PDMS is a widely used polymer, due to its flexibility, manufacturing ease, transparency, biocompatibility and super-hydrophobicity. It is also widely used as triboelectric material in the construction of TENGs for a broad range of applications \cite{Zhu2016,Ko2014,Zheng2016}.
On the other hand, Nylon has good mechanical properties (strength and stiffness), high impact resistance, is easy to fabricate and maintains its properties over a large temperature range \cite{Goetz2003}.

Herein, we study the effect of the variation of the thickness of the PDMS triboelectric layer and of the surface area on the generated electrical outputs when PDMS and Nylon come into contact.
We aim to best understand and enhance the triboelectric effect and consequently to optimize future prototypes.

\section{Experimental details}
In this study we used a SYLGARD 182 Silicone Elastomer kit, supplied in two parts
consisting of the PDMS base and the curing agent component (Dimethyl,Methylhydrogen Siloxane).
The base and the curing agent were mixed using a weight ratio of 10:1.
We used the spin-coating technique at different rotation-velocities to deposit the fabricated PDMS on an aluminium substrate which will serve as one of the electrodes.
The rotation velocity of the spinner was variated from 500 to 5000 rpm, leading to PDMS thicknesses from 13 to 220 $\mu$m \cite{Mata2005,Koschwanez2009,All,Mata2005,Koschwanez2009,Elveflow2013}.
Curing of PDMS was performed at 80 $^{\circ}$C for two hours in an oven.
The other triboelectric material (Nylon) was used in thin film form with a thickness of 50 $\mu$m.

To measure the generated electrical signals, an aluminium tape was attached to both triboelectric materials to serve as electrode and placed into acrylic plates (with 20 cm$^{2}$).
We then used a home-made systematic testing system that makes the two triboelectrics materials come into contact.
Measurements of the generated current, voltage and power as a function of the load resistances ($R_{L}$) were then performed using a circuit board with resistors from 100 to 1 G$\Omega$.

\section{Results and discussion}
\subsection{Thickness of the Triboelectric Layer}
\begin{figure*}[!ht]
\centering
\includegraphics[height=4.5in]{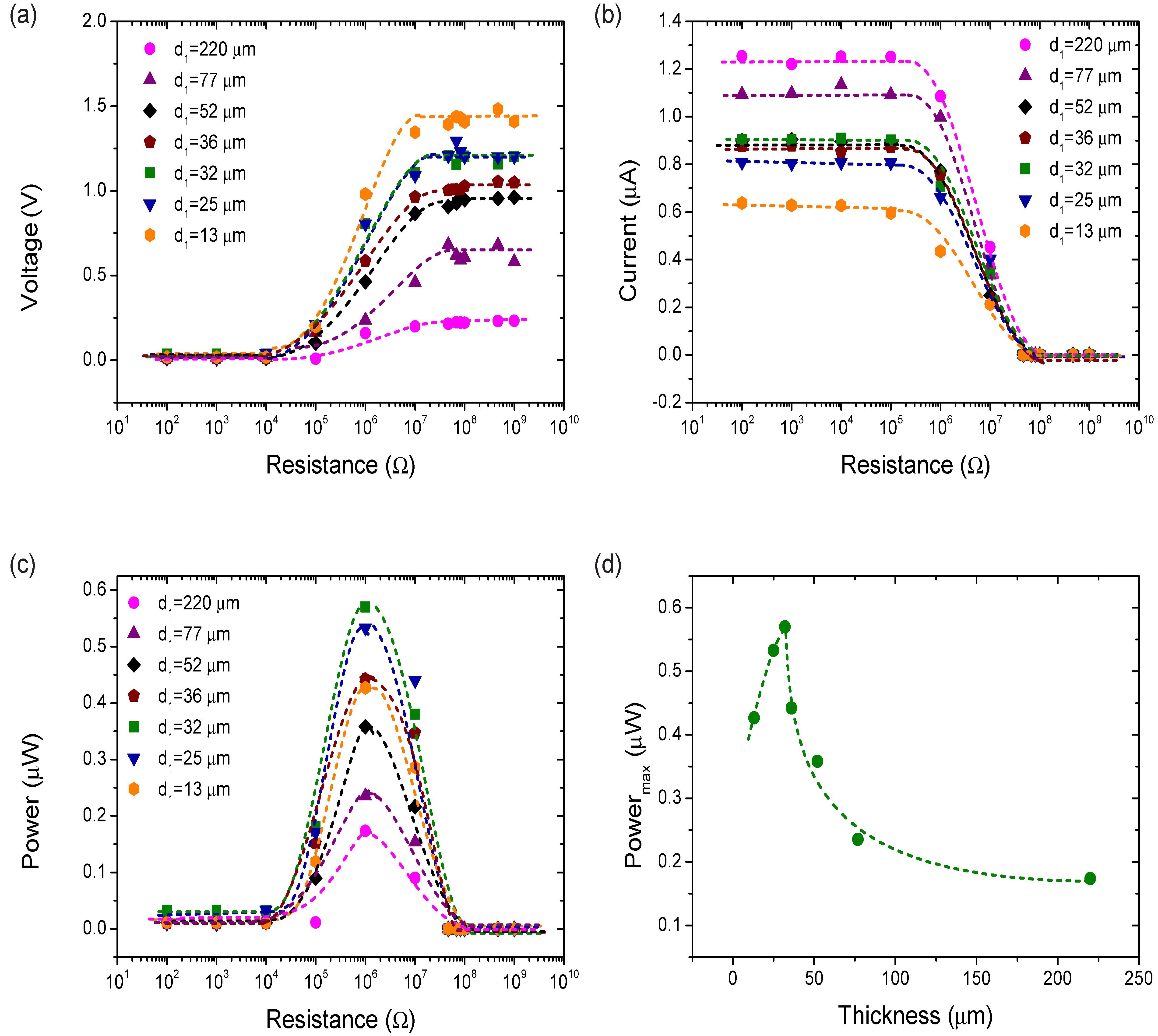}
\caption[]{\small{Influence of the PDMS triboelectric layer thickness on the output (a) voltage, (b) current and (c) power generated as a function of the load resistance. (d) Maximum output power generated for the contact-separation mode of different PDMS and Nylon plates.}}
\label{fig:2}
\end{figure*}

Aiming the improvement of the triboelectric effect, we studied the influence of the PDMS layer thickness on the generated electrical outputs.
PDMS was the triboelectric material chosen to vary the thickness (between 13 and 220 $\mu$m), while Nylon was used in film form [Fig. \ref{fig:2}(a)].
When the different samples of PDMS come into contact with the Nylon plate, we systematically measured the voltage, current and corresponding power.

\begin{figure*}[!ht]
\centering
\includegraphics[height=4.5in]{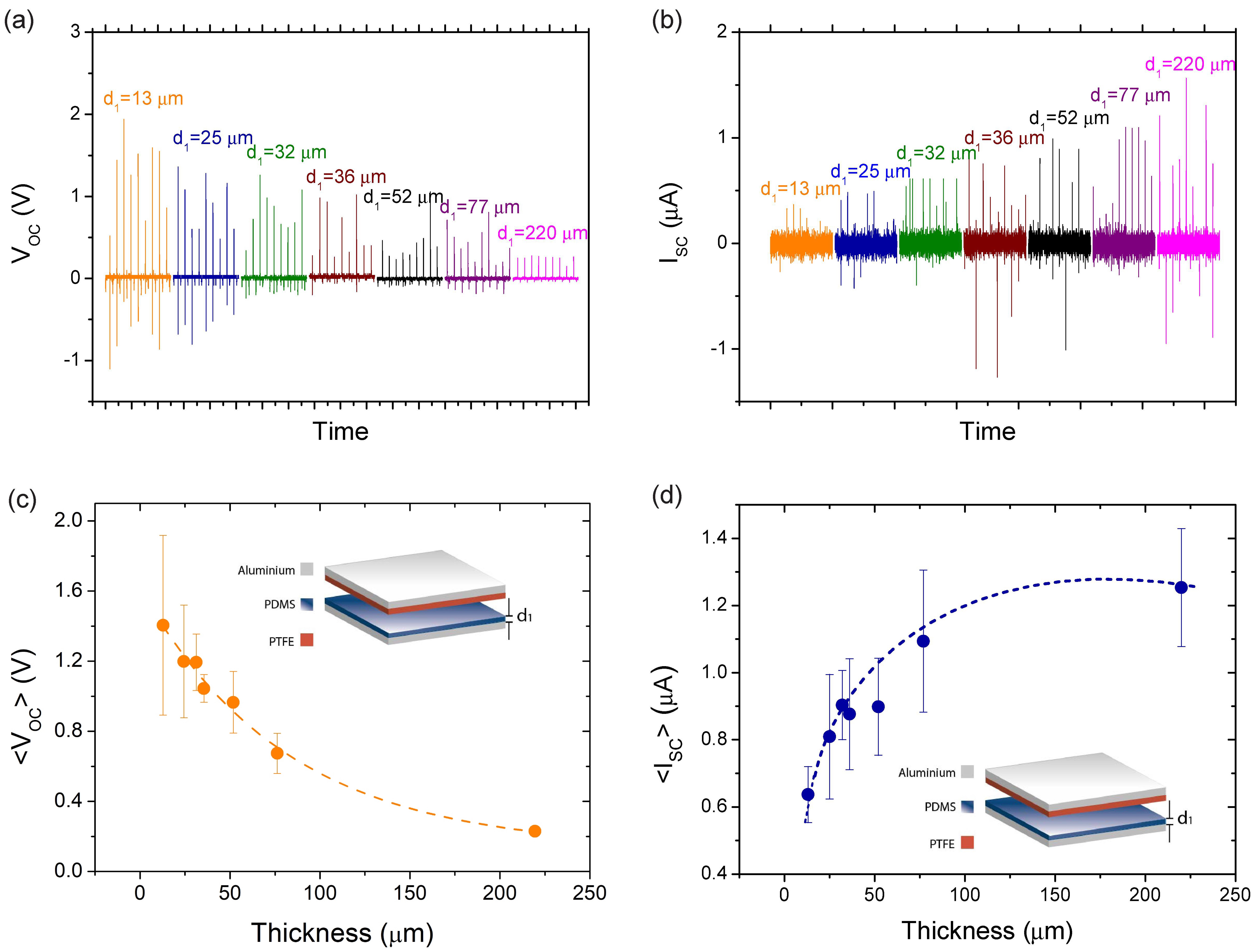}
\caption[]{\small{(a) Open-circuit voltage and (b) short-circuit current peaks, (c) mean open-circuit voltage and (d) mean short-circuit current for the samples with different PDMS layer thickness.}}
\label{fig:1}
\end{figure*}

Figures \ref{fig:1}(a) and (b) show the measured open-circuit voltage and short-circuit current, respectively, for different PDMS layer thicknesses.
From these graphs it is possible to observe that the voltage (current) decreases (increases) with increasing thickness of the PDMS triboelectric layer.
Taking the average values of the open-circuit voltage and short-circuit current peak for all PDMS thicknesses we obtained the graphs shown in Figs. \ref{fig:1}(c) and (d), respectively.
A maximum voltage of 1.4 V for 13 $\mu$m of PDMS was obtained.
On the other hand, the minimum voltage (0.2 V) occurred for a PDMS thickness of 220 $\mu$m.
The decrease of $<V_{OC}>$ with increasing thickness is apparently not in agreement with present TENG theories \cite{Niu2013,Wang2016}.
It must then be related with intrinsic properties of the PDMS films, such as stiffness , hardness or roughness and their thickness dependence.
The current has the opposite behaviour since, as the thickness increases, the current also increases, saturating at a value of $\sim$ 1.3 $\mu$A [Fig. \ref{fig:2}(b)].
These results are in accordance with the theory of triboelectric nanogenerators [Eq. (3)].

We then measured the voltage and current from 100 to 1 G$\Omega$ , as seen in Figs. \ref{fig:2}(a) and (b).
Figure \ref{fig:2}(a) illustrates the generated voltage and, for all samples, the voltage increased above a load resistance of 10 k$\Omega$, tending to a constant value (open-circuit voltage).
We also measured the current flowing in the circuit for the same range of resistances and noticed the opposite behaviour since, with increasing resistance, the current values decrease [Fig. \ref{fig:2}(b)].
These results confirm that increasing the PDMS thickness leads to a decrease (increase) of the generated voltage (current).
On the other hand, the maximum output power occurred for a PDMS layer thickness of 32 $\mu$m and has a value of 0.56 $\mu$W [which corresponds to a power density of 0.55 mW/m$^{2}$; Fig. \ref{fig:2}(c)].
Figure \ref{fig:2}(d) shows the maximum generated power as a function of the PDMS thickness layer.
It is possible to verify two different behaviours: up to 32 $\mu$m of PDMS, the generated power increases with increasing thickness (reaching a maximum value of 0.56 $\mu$W); from 32 to 220 $\mu$m the generated power decreases sharply, reaching a minimum value of 0.23 $\mu$W (which corresponds to a power density of 0.23 mW/m$^{2}$) .

\subsection{Area of the Triboelectric Surfaces}
To investigate the relationship between the electric outputs and the area of the triboelectric surfaces, a set of systematic measurements was also performed.
In this study, we changed the areas of the triboelectric materials between 2.5 and 15 cm$^{2}$.
Similar to the previous study, we used a PDMS-Nylon tribo-pair, although in this study the thickness of the PDMS remained constant (= 32 $\mu$m).

\begin{figure*}[!ht]
\centering
\includegraphics[height=4.5in]{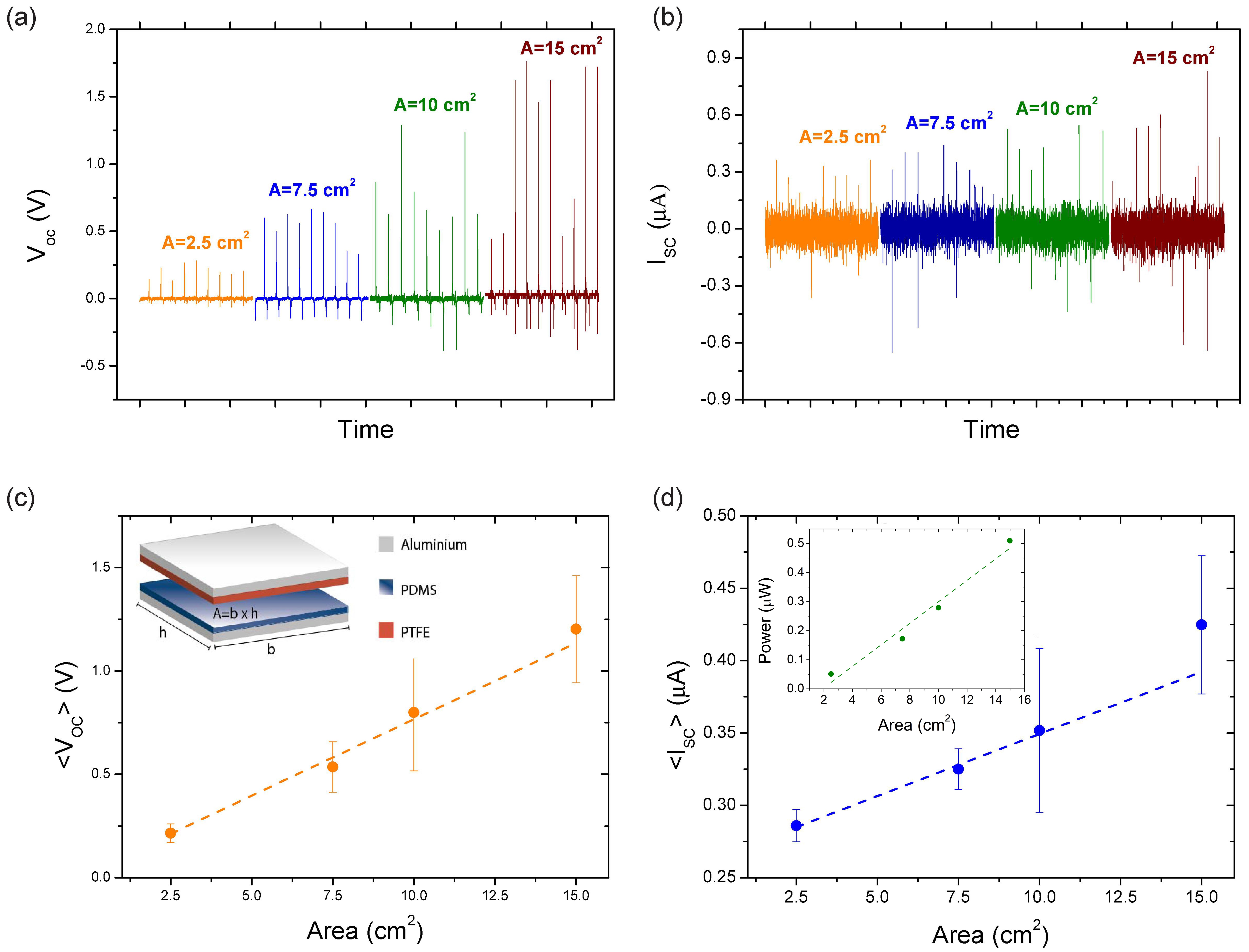}
\caption[]{\small{(a) Open-circuit voltage, (b) short-circuit current, (c) mean open-circuit voltage and (d) mean short-circuit current obtained for different areas of the triboelectric surfaces. The inset is the area dependence of the maximum power.}}
\label{fig:3}
\end{figure*}

In Fig. \ref{fig:3}(a), it is possible to observe the open-circuit voltage for the four different values of triboelectric surface area, clearly demonstrating an increase of the voltage peaks with increasing contact area.
In Fig. \ref{fig:3}(b) are represented the short-circuit currents obtained for the same triboelectric surface areas.
The output current also increases with the increase of the contact area, due to the increased amount of transferred charges.
Similarly, we proceeded to calculate the average of the voltage and current peak values for the different areas of the triboelectric surfaces [Figs. \ref{fig:3}(c) and (d)].
Figure \ref{fig:3}(c) shows the values of the mean open-circuit voltage generated for the Nylon-PDMS pair in the contact-separation mode.
The maximum value of the generated voltage was 1.2 V and occurred for a surface area of 15 cm$^{2}$.
The maximum generated current was 0.43 $\mu$A for a contact area of 15 cm$^{2}$ [Fig. \ref{fig:3}(d)].
Thus, we observed that the increase of the macroscopic contact area leads to the increase of the output voltage and current.

We again measured the electric outputs for the four different areas when connected to variable load resistances (from 100 to 1 G$\Omega$; not shown).
As expected, the voltage has maximum values (1.2 V) at high resistances, whereas the current reaches maximum values (0.43 $\mu$A) for small resistances.
The area dependence of the maximum power is shown in the inset of Fig. \ref{fig:3}(d), displaying a linearly increasing tendency with the increase of sample area and reaching an output power of 0.5 $\mu$W (corresponding to a power density of 0.33 mW/m$^{2}$).

\section{Conclusion}
In summary, we studied the influence of the thickness of the triboelectric layer and of the contact area of the triboelectric surfaces on the triboelectric effect.
By changing the PDMS thickness layer, we obtained the relationship between the PDMS thickness layer and the electrical output, when PDMS comes into contact with Nylon.
With increasing PDMS thickness, the voltage values decreased but the current increased.
The maximum output power occurred for a PDMS layer thickness of 32 $\mu$m and had a value of 0.56 $\mu$W.
We also observed that increasing the contact surface of the triboelectric materials increases the electrical output generated for the contact-separation of the tribo-pair PDMS-Nylon.
For a surface area of 15 cm$^{2}$, we achieved an output power density of 0.33 mW/m$^{2}$.

\section*{Acknowledgments}
The authors acknowledge funding from FEDER and ON2 through project UTAP-EXPL/NTec/0021/2017 and from FCT through the Associated Laboratory – Institute of Nanoscience and Nanotechnology (POCI-01-0145-FEDER-016623). J. V. acknowledges financial support through FSE/POPH and project UTAP-EXPL/NTec/0021/2017. C. R. thanks inanoEnergy for the research grant.


\bibliographystyle{model3-num-names}
\bibliography{bibliography}

\end{document}